\documentclass[conference]{IEEEtran}
\IEEEoverridecommandlockouts
\usepackage{cite}
\usepackage{amsmath,amssymb,amsfonts}
\usepackage{algorithmic}
\usepackage{graphicx}
\usepackage{textcomp}
\usepackage{xcolor}

\usepackage{numprint}
\npdecimalsign{.}
\newcommand{\eer}[1]{\nprounddigits{2}\numprint{#1}}

\usepackage[colorlinks=true, allcolors=blue]{hyperref}

\def\BibTeX{{\rm B\kern-.05em{\sc i\kern-.025em b}\kern-.08em
    T\kern-.1667em\lower.7ex\hbox{E}\kern-.125emX}}
\begin{document}

\title{Probing the Information Encoded in Neural-based Acoustic Models of Automatic Speech Recognition Systems}

\author{\IEEEauthorblockN{Quentin Raymondaud}
\IEEEauthorblockA{
\textit{LIA - Avignon University}\\
Avignon, France \\
quentin.raymondaud@univ-avignon.fr}
\and
\IEEEauthorblockN{Mickael Rouvier}
\IEEEauthorblockA{
\textit{LIA - Avignon University}\\
Avignon, France \\
mickael.rouvier@univ-avignon.fr}
\and
\IEEEauthorblockN{Richard Dufour}
\IEEEauthorblockA{
\textit{LS2N - Nantes University}\\
Nantes, France \\
richard.dufour@univ-nantes.fr}
}

\maketitle

\begin{abstract}
Deep learning architectures have made significant progress in terms of performance in many research areas. The automatic speech recognition (ASR) field has thus benefited from these scientific and technological advances, particularly for acoustic modeling, now integrating deep neural network architectures. However, these performance gains have translated into increased complexity regarding the information learned and conveyed through these ``black-box'' architectures. Following many researches in neural networks interpretability, we propose in this article a protocol that aims to determine which and where information is located in an ASR acoustic model (AM). To do so, we propose to evaluate AM performance on a determined set of tasks using intermediate representations (here, at different layer levels). Regarding the performance variation and targeted tasks, we can emit hypothesis about which information is enhanced or perturbed at different architecture steps. Experiments are performed on both speaker verification, acoustic environment classification, gender classification, tempo-distortion detection systems and speech sentiment/emotion identification. Analysis showed that neural-based AMs hold heterogeneous information that seems surprisingly uncorrelated with phoneme recognition, such as emotion, sentiment or speaker identity. The low-level hidden layers globally appears useful for the structuring of information while the upper ones would tend to delete useless information for phoneme recognition.
\end{abstract}

\begin{IEEEkeywords}
automatic speech recognition, acoustic model, deep neural network, artificial intelligence explanation
\end{IEEEkeywords}

\section{Introduction}
\label{intro}

In recent years, automatic speech recognition (ASR) systems have experienced major advances, which have resulted in very significant performance gains. These upheavals have been particularly visible with the integration of deep learning approaches at both acoustic and linguistic levels, in particular through neural network architectures, coupled with the use of massive speech data.

In a classical ASR framework, an acoustic model (AM) is trained to recognize phonemes. Acoustic modeling then went from Hidden Markov Models (HMMs) with Gaussian Mixture Models (GMMs)~\cite{Jelinek76,su2010gmm} to Deep Neural Network (DNN) approaches~\cite{bhatt2020acoustic,dighe2020quantifying}. Even if significant gains have been observed, it is still difficult to understand and report information learned by these DNN methods. This need for explainability is found in many areas, whether in image~\cite{singh2020explainable}, text~\cite{karim2021deephateexplainer} or speech~\cite{bharadhwaj2018layer} processing.

While many works focused on improving AMs, relatively few sought to shed light on the information they conveyed. In speech recognition, only a small number of papers~\cite{mohamed12,nagamine2017understanding} study representations of speech learned by feed-forward DNNs used in acoustic modeling for the phone recognition task. Other works may be found in end-to-end ASR systems, that directly encode acoustic and linguistic information in a single deep learning architecture, that identify the information encoded through the final model, but also the different layers that compose it. Classical ASR systems, comprising separate acoustic and linguistic models, are however still widely used.

Works mainly seek to understand where the information is located in the different neuronal layers, at the time being simply at the phone~\cite{belinkov2017analyzing}, grapheme level~\cite{belinkov2019analyzing} and speaker representation~\cite{williams2019disentangling,Raj_2019}. Another interesting work~\cite{li2020does} proposed to analyze each layer of an end-to-end ASR system by synthesizing speech from hidden representations, and then to evaluate by listening to the synthesized speech. They observed that the lowest layer shows less speaker variability and noise. One limit of these works is that they are focused on analyzing a single piece of information contained in these networks, even though the information contained in the AMs can be broad (paralinguistic~\cite{kumar2021towards}, speaker~\cite{mdhaffar2021retrieving}, etc.).

In this paper, we propose to study and quantify the information contained in the acoustic models trained within the framework of a state-of-the-art classical ASR system. Following previous studies, we seek to analyze the different layers composing an AM based on a factorized time-delay neural network (TDNN-F) architecture~\cite{povey2018semi}. Unlike other works, a set of tasks have been identified, making it possible to highlight information of different natures and its dynamics of appearance (and disappearance) in the networks in the context of phoneme recognition. We are then trying to determine whether some information are removed, or still preserved, and at which layers level. For each targeted task, we suppose that specific information snippets are important. 
We therefore seek to answer the question of whether the AM forms a generalist pseudo-uniform vector space in despite of a very specialized phoneme-oriented training process. In addition, we freely released the recipes to facilitate further research$\footnote{\url{https://gitlab.com/Raymondaud.Q/probing-acoustic-model}}$.

The paper is organized as follows. Section~\ref{sec:asr} summarizes the studied acoustic model used in the context of automatic speech recognition. Then, Section~\ref{sec:probing} presents the
 protocol that we used for highlighting the presence (or absence) of information of different nature in the AM. In Section~\ref{sec:experimental_protocol}, we describe the different probing tasks and associated datasets. A qualitative analysis is proposed in Section~\ref{sec:exp_results} before concluding and giving perspectives in Section~\ref{sec:conclusions}. 

\section{Acoustic model architecture}
\label{sec:asr}

The acoustic model is one of the main modules of a classical ASR system. It makes possible the recognition of basic speech units, here the phonemes, and has therefore been trained to classify them in a targeted language given an acoustic signal. One of the difficulties of this acoustic modeling lies in the fact that the signal conveys a lot of information other than the phonemes themselves: linguistic, noise, speaker, etc. This information is often encoded in such a complex manner that the signal exhibits a great deal of variability, which makes the phoneme classification task difficult.

AM accuracy has recently been improved using DNNs that can handle highly correlated features. Hidden layers have then the ability to de-correlate the useless information (environment, speaker, etc.) in order to focus on the useful one for a targeted final task. DNNs allowed us to significantly improve systems, but, at the same time, made our knowledge related to the information contained in an AM more complex.

The acoustic model studied in this paper is a DNN based on the TDNN-F architecture~\cite{povey2018semi}. TDNN-F has the same structure as TDNN but the layers of the former are compressed by singular value decomposition and trained by random initialization with semi-orthogonal constraints in one of the two factors of each matrix. TDNN-F is constructed with 16 hidden layers, each containing 1,536 hidden nodes. TDNN-F model was trained on the Librispeech dataset~\cite{panayotov2015librispeech} using the Kaldi toolkit~\cite{povey2011kaldi}. The \texttt{s5} recipe$\footnote{\url{https://github.com/kaldi-asr/kaldi/blob/master/egs/librispeech/s5/}}$ has been followed. Note that no speaker adaptation method has been applied in order to keep a genericity at the speaker level.

\section{Proposed protocol}
\label{sec:probing}

\begin{figure}[h]
    \includegraphics[width=0.49\textwidth]{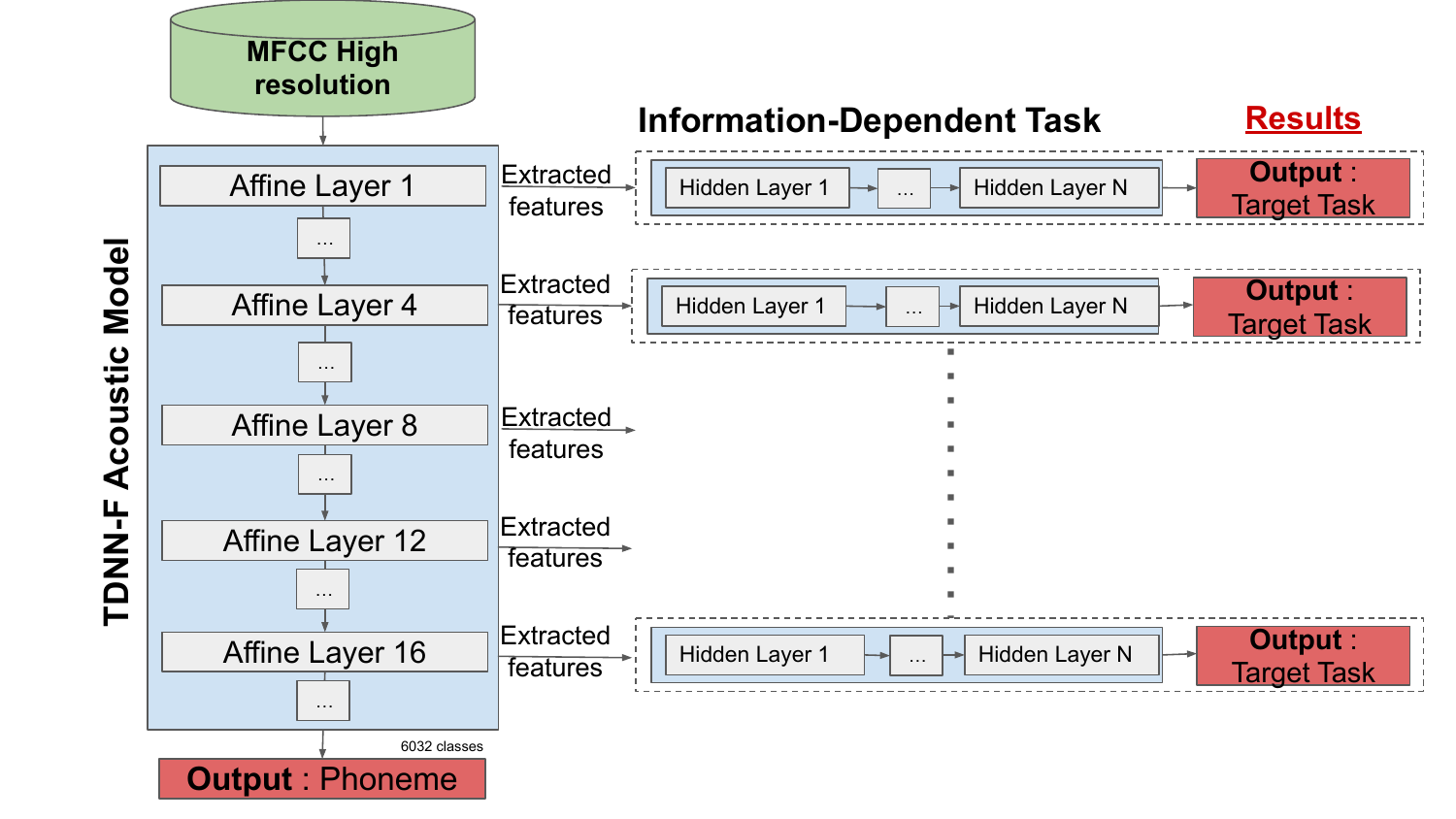}
    \caption{\label{fig:architecture}Proposed protocol for acoustic model information probing.}
\end{figure}

We make the assumption that specific information is contained in the hidden layers and may change depending on their level in the neural architecture. We propose to probe the presence (or absence) of specific information in the AM using several targeted classification tasks carried out with parameters extracted from several hidden layer. Our goal is to reveal the link between the features given by the AM hidden layers and the tasks. The classification tasks are carried out, each time with parameters extracted from specific hidden layers of the AM. High performance should then reveal important task-related characteristics contained in these layers, and vice versa. Figure~\ref{fig:architecture} summarizes the protocol of our approach based on the TDNN-F acoustic model architecture (see Section~\ref{sec:asr}).


For every classification tasks, we use a common ECAPA-TDNN classifier~\cite{desplanques2020ecapa}. The ECAPA-TDNN architecture uses cutting-edge techniques: Multilayer Feature Aggregation (MFA), Squeeze-Excitation (SE) and residual blocks. This model has recently shown impressive performance in the speaker verification~\cite{thienpondt2021idlab}.

ECAPA-TDNN had the following parameters: the number of SE-Res2Net Blocks is set to 4 with dilation values 2, 3 and 4 to blocks; the number of filters in the convolutional frame layers C is set to 1,024 equal to the number of filter in the bottleneck of the SE-Res2Net Block; embedding layer size is set to 192. For training, we used one cycle learning rate scheduler with SGD or ADAMW optimizer depending on tasks.

\begin{table*}[h!]

\center

\caption{\label{tbl:results}Performance obtained for each probing task at different TDNN-F layer levels as well as the MFCC baseline. The scores given in \textbf{bold} and in \underline{underline} are respectively the highest and lowest ones. Note that scores are provided in terms of accuracy for all tasks except for speaker verification (EER).}

\resizebox{1.8\columnwidth}{!}{
\begin{tabular}[|c]{|l||c|c|c|c|c|c|}
\hline

 & {\bf Speaker} & {\bf Speaking} & {\bf Speaker} & {\bf Acoustic} & {\bf Speech} & {\bf Speech} \\
& {\bf Gender} & {\bf Rate} & {\bf Verification} & {\bf Environments} & {\bf Sentiment} & {\bf Emotion}  \\

\hline\hline

MFCC & \eer{0.9635217176059825} & \underline{\eer{0.6487597383986035}} & \textbf{\eer{1.967}} & \underline{\eer{0.6847}} &  \underline{\eer{0.42161747796090454}} & \underline{\eer{0.22920659256420084}} \\\hline

Layer1 & \eer{0.95} & \eer{0.65} & \eer{2.134} & \eer{0.688} & \eer{0.4883096972019931} & \eer{0.33652740513606744} \\

Layer2 & \eer{0.95} & \eer{0.68} & \eer{2.189} & \eer{0.7436} & \textbf{\eer{0.5021080873898045}} & \textbf{\eer{0.4814105021080874}}\\

Layer4 & \eer{0.9435889418692515} & \eer{0.68} & \eer{2.325} & \textbf{\eer{0.7576}} & \eer{0.48715983135300883} & \textbf{\eer{0.4814105021080874}}\\

Layer6 & \eer{0.95} & \eer{0.68} & \eer{2.573} & \eer{0.7551} & \eer{0.4672288233039479} & \eer{0.46799540053660404} \\

Layer8 & \textbf{\eer{0.9700161546679174}} & \eer{0.68} & \eer{2.841} & \eer{0.7537} & \eer{0.4645458029896512} & \textbf{\eer{0.4810272134917593}} \\

Layer10 & \eer{0.9502332004481617} & \textbf{\eer{0.70}} & \eer{3.374} & \eer{0.7451} & \eer{0.47566117286316595} & \textbf{\eer{0.4814105021080874}}\\

Layer12 & \eer{0.9502332004481617} & \eer{0.68} & \eer{3.796} & \eer{0.744} & \eer{0.4591797623610579} & \eer{0.35032579532387886} \\

Layer14 & \eer{0.92} & \eer{0.67} & \eer{4.343} & \eer{0.7238} & \eer{0.46032962821004214} & \eer{0.24032196243771559}\\

Layer16 & \underline{\eer{0.8796802939107324}} & \eer{0.66} & \underline{\eer{5.089}} & \eer{0.7143} & \eer{0.4695285550019164} & \textbf{\eer{0.4814105021080874}} \\\hline 
\end{tabular}
}
\end{table*}



\section{Probing tasks}
\label{sec:experimental_protocol}

This section describes the 5 different tasks studied for probing information contained in neural-based AMs: speaker verification, speaking rate, speaker gender, channel-related information (acoustic environments) and paralinguistic information (speech sentiment / emotion).

\subsection{Speaker verification}

Speaker verification refers to the task of verifying the identity claimed by a speaker from that person's voice. This task measures how well the acoustic model encodes the speaker information (speaker-specific traits), which is crucial for the speaker verification task. We extract, for the claimed identity and from the speaker person's voice, a speaker embedding and verify the matching thanks to cosine similarity. Speaker embedding is a high-level speaker representation extracted directly from its acoustic excerpts from ECAPA-TDNN.
Equal Error Rate (EER) is used as the performance criterion of speaker verification task (threshold value such that false acceptance and miss rates are equals).

The system has been trained on the VoxCeleb2 dataset~\cite{chung2018voxceleb2}, only on the development partition, which contains speech from 5,994 speakers. Systems are evaluated on Voxceleb1-E Cleaned~\cite{nagrani2017voxceleb} dataset (which involves 579,818 trials). Note that the development set of VoxCeleb2 is completely disjoint from the VoxCeleb1-E Cleaned dataset ({\it i.e.} no speaker in common).


\subsection{Speaking rate}

In this task, we augment all utterances by 3-way speed perturbation with rates of 0.85, 1.0 and 1.15. This task measures whether the acoustic model can capture information on speaking rate. We train a three-classes ECAPA-TDNN classifier and report the classification accuracy. The system has been trained on Voxceleb2 dataset and evaluated on Voxceleb1 dataset (which contains 153,516 utterances).

\subsection{Speaker gender}

This task measures whether the acoustic model can distinguish between gender ({\it i.e.} male or female). We train a two-classes ECAPA-TDNN classifier and report the classification accuracy. The system has been trained on Voxceleb2 dataset and evaluated on Voxceleb1 dataset (which contains 153,516 utterances). We note that the datasets are fairly gender balanced, (55\% male and 45\% female).

\subsection{Acoustic environments}

This task measures whether the acoustic model captures information on acoustic environments (air\_conditioner, car\_horn, children\_playing, dog\_bark, drilling, enginge\_idling, gun\_shot, jackhammer, siren and street\_music). We train a ten-classes ECAPA-TDNN classifier and report the classification accuracy. The system has been trained using the UrbanSound8k dataset~\cite{Salamon:UrbanSound:ACMMM:14} that contains 8,732 sounds divided into 10 classes. We used the recipe proposed in Speechbrain~\cite{ravanelli2021speechbrain} that split the corpus in 10-fold cross validation.

\subsection{Speech sentiment/emotion recognition}

Speech Sentiment Recognition (SSR) and Speech Emotion Recognition (SER) aim to classify speech records respectively in three-classes (positive, neutral and negative) and seven-classes (anger, disgust, fear, joy, neutral, sadness and surprise). We report the classification in terms of accuracy.

The system has been train on Multimodal EmotionLines Dataset (MELD) corpus~\cite{poria2018meld}. This corpus is composed of 13,000 utterances from Friends TV (sitcom). Each utterance is annotated in terms of sentiment and emotion.

\section{Experiments and Results}
\label{sec:exp_results}

Table~\ref{tbl:results} summarizes the performance obtained by the different targeted probing tasks. Scores are expressed in terms of EER for speaker verification, and accuracy for all the others. We compare the performance obtained for each task at different TDNN-F layer levels (from Layer1 to Layer16) as well as the MFCC (Mel-frequency cepstral coefficients) baseline. Note that MFCCs are the acoustic features used as input in the TDNN-F. 

Globally, the vector representations from the hidden layers provide better classification results than a conventional representation of parameters extracted from a speech signal (here, MFCCs). Only the speaker verification task performs better with MFCCs: the AM therefore tends to suppress information linked to the speaker identity, that must participate negatively in the phoneme recognition task, contrary to the other tasks. We observe the same phenomenon in Self-Supervised Learning models such as wav2vec2. In~\cite{chen2022large}, the authors observe that wav2vec2 models tend to suppress information linked to the speakers identity on the upper layers. This tends to confirm that information linked to the speakers identity is not useful for phoneme identification task and must be suppressed.

It therefore appears that the hidden layers contain heterogeneous and structured information from the speech signal, whether at the speaker, acoustic environment or the paralinguistic level. Depending on their depth ({\it i.e.} the position of the hidden layer in the network), they do not seem to encode the same information. Lower levels pick up surrounding noise better, with best performance achieved with Layer4 on the acoustic environments task (accuracy of 0.76). In the same way, speech sentiment and speech emotion quickly reach best performance at Layer2, with a respective accuracy of 0.50 and 0.48. Speaker gender and speaking rate achieve the highest performance at the middle level of the network with Layer8 (accuracy of 0.97) and Layer10 (accuracy of 0.70) respectively. Concerning the speaker gender classification, by listening the utterances misclassified by the system, we realize that the system has classified the speakers in terms of low and high voice (low and high voice are respectively assigned to male and female). This observation tends to confirm that information on high and low voice is important to help the phoneme classification task.

However, note that while performance tends to decrease in the highest hidden layers, these remain at a very good level of accuracy for most of the tasks. It therefore seems that as the network is built, the lower hidden layers structure the information, while the higher hidden layers will suppress information harmful to phoneme recognition. It is then interesting to note that information which seems useless in AM for the ASR task is preserved.

\section{Conclusion and Perspectives}
\label{sec:conclusions}


In this paper, we proposed a protocol that aims at highlighting information contained in acoustic models used in ASR systems. To do so, we finely studied a neural-based AM using different speech-oriented tasks. By analyzing the performance obtained by each specific task in the different hidden layers of the studied TDNN-F acoustic model, we were able to realize the information contained at various levels of the AM, whether at the speaker, at the acoustic environment, or at the paralinguistic information related to emotion and sentiment.

Our study showed that TDNN-F AMs provide information that encode gender, speaking rate, speaker identity, emotion and sentiment-related information inside this neural-based model, even though these AMs have been trained to recognize phonemes. The proposed work highlighted that the information is not encoded in the same way within the AM: low-level layers tend to structure information, with a continuous increase in performance, until tending towards a suppression of information, which is observed by a final drop in results on our targeted tasks.


In future works, we will endeavor to expand the tasks (accent, age, etc.) to obtain additional information about which information the AM encodes from the MFCCs. Similarly, we wish to focus on other representations of the acoustic signal, in particular the wav2vec unsupervised representation.


\section{Acknowledgements}

This work was financially supported by the DIETS project financed by the French National Research Agency (ANR) under contract ANR-20-CE23-0005 and was granted access to the HPC resources of IDRIS under the allocation 2021-A0111012991 made by GENCI.

\bibliographystyle{IEEEtran}
\bibliography{custom}

\end{document}